\documentclass[twocolumn,showpacs,pra,amsmath,amssymb]{revtex4}
\usepackage{graphicx,amsmath}

\begin{document}

\title{Semiconductor-based Geometrical Quantum Gates}
\author{
Paolo Solinas,$^{*}$ Paolo Zanardi,$^{\dag}$ Nino Zangh\`{\i},$^{*}$
and Fausto Rossi$^{\dag,\ddag}$
}
\affiliation{
$^*$ Istituto Nazionale di Fisica Nucleare (INFN) and
Dipartimento di Fisica, Universit\`a di Genova,
Via Dodecaneso 33, 16146 Genova, Italy \\
$^\dag$ Institute for Scientific Interchange (ISI),
Viale Settimio Severo 65, 10133 Torino, Italy \\
$^\ddag$ Istituto Nazionale per la Fisica della Materia (INFM) and
Dipartimento di Fisica, Politecnico di Torino, Corso Duca degli
Abruzzi 24, 10129 Torino, Italy }

\date{\today}

\begin{abstract}
 We propose an implementation scheme for holonomic, i.e., geometrical, quantum information processing based on semiconductor  nanostructures.
Our quantum hardware consists of coupled semiconductor macroatoms addressed/controlled by ultrafast multicolor laser-pulse sequences. 
More specifically, logical qubits are encoded in excitonic states with different spin
polarizations and manipulated by  adiabatic time-control  of the
laser amplitudes . The two-qubit gate is realized in a geometric fashion
by exploiting dipole-dipole coupling between excitons in neighboring
quantum dots.
\end{abstract}

\maketitle

In the last few years the promise to outperform classical protocols for  information manipulation has attracted a huge  interest in Quantum Information
Processing (QIP)\cite{1}.
Unfortunately, processors working according to the rules of quantum mechanics are, even in principle, extremely delicate objects:
On the one hand the  unavoidable coupling with
uncontrollable degrees of freedom (the environment) spoils the unitary nature of the dynamical evolution, i.e., decoherence.
On the other hand, extreme capabilities in quantum-state control are required;
indeed, typically even very small manipulation imperfections will eventually drive the processing system into a   ``wrong'' output state.
It is therefore clear that any general strategy that appears to be able to cope with  this sort of inherent fragility  of QIP
is worthwhile of serious consideration.

So far, quantum error-correction \cite{2}, error-avoiding \cite{3}, and
error-suppression techniques~\cite{4},\cite{5} have been developed at the theoretical level.
They are mainly devoted to stabilize quantum information against computational errors induced by  coupling with the environment, 
and are based on either  the idea of hiding information to the detrimental effects of noise or to dynamically get rid of the noise itself.
 All of these strategies  require extra-physical resources in terms of either  qubits or additional manipulations.

A further, conceptually fascinating, strategy for  the stabilization of quantum information is provided by the  ``topological approach'' \cite{6},\cite{7}.
In such QIP schemes gate operations depend just on topological ---i.e., global--- features of the control process, and are therefore largely insensitive to
local inaccuracies and fluctuations. This approach can be regarded as a sort of
``digitalization'' of a continuous dynamical system and it allows  in principle
a very appealing liberty in the control process to be implemented.

As a matter of fact, such topological schemes are so far pretty abstract: information has to be encoded in highly non-local quantum states of many-body systems interacting in an exotic fashion.
A significant intermediate step in this direction is given by the so-called ``Holonomic'' Quantum Computation (HQC) \cite{8}, \cite{9}.
In this framework  quantum information is encoded in a $n$-fold degenerate eigenspace of a
family of quantum Hamiltonians depending on dinamically controllable parameters $\lambda$.
Quantum gates are enacted by driving  the $\lambda$' s along suitable loops $\gamma$ within the manifold. The non-trivial dependence of Hamiltonian eigenvectors
on the $\lambda$ results in non-trivial transformations
of the initially prepared state. Such transformations ---known as {\em holonomies}---
generalize to the non-Abelian  case the celebrated Berry's phase \cite{10}.
When the loops are undergone in an adiabatic way  holonomies can be
explicitly computed in terms of the Wilczek-Zee gauge connection \cite{11},
and conditions for achieving universality are simply stated \cite{8}.

As for the topological schemes, the built-in fault-tolerant features of the holonomic approach are related to the fact that {\em the holonomies depend on some global geometrical feature, e.g., area, of the $\gamma$, and not on the way the loops are actually  realized}.

Quantum gates based on (Abelian) Berry phases  have been experimentally realized using
nuclear magnetic resonance (NMR) schemes \cite{12}, and recently proposed for mesoscopic Josephson junctions~\cite{13} and anyonic excitations in Bose-Einstein
condensates \cite{14}.
Non-adiabatic realizations of Berry's phase logic gates have been studied
as well \cite{15}, \cite{16}
More recently, schemes for the experimental implementation of  non-Abelian
holonomic gates have been proposed for atomic physics \cite{17},
ion traps \cite{18}, Josephson junctions \cite{19}, Bose-Einstein 
condensates \cite{20}, and neutral atoms in cavity \cite{21}.

We propose the first implementation scheme for the realization of
an universal set~\cite{22} of non-Abelian holonomic quantum gates in semiconductor nanostructures \cite{23}.
As we shall see, in the proposed strategy a central role is played by the holonomic structure  introduced in \cite{17} and \cite{18} as well as by the
exciton-exciton interaction mechanism  exploited in the all-optical semiconductor-based QIP scheme proposed in \cite{24}.
The proposed quantum hardware is given by an array of semiconductor quantum dots (QDs) \cite{25}, often referred to as macroatoms;
 our computational degrees of freedom are interband optical excitations, also called excitonic transitions.
Indeed, an exciton is a Coulomb-correlated electron-hole pair produced by promoting an electron from the valence band with total angular momentum $J_{tot}=3/2$ to
the conduction band with $J_{tot}=1/2$.
For a GaAs-based quantum-dot structure, the confining potential along the growth ($z$) direction breaks the symmetry and lifts the degeneracy in the 
valence band \cite{23},\cite{25};
the states ($| J_{tot}, J_z \rangle$) of the quadruplet $J_{tot}=3/2$ are then energetically separated into $J_z=\pm 3/2$
---heavy holes (HH)--- and $J_z=\pm 1/2$ ---light holes (LH)---.

\begin{figure}[t]
  \includegraphics[height=12cm]{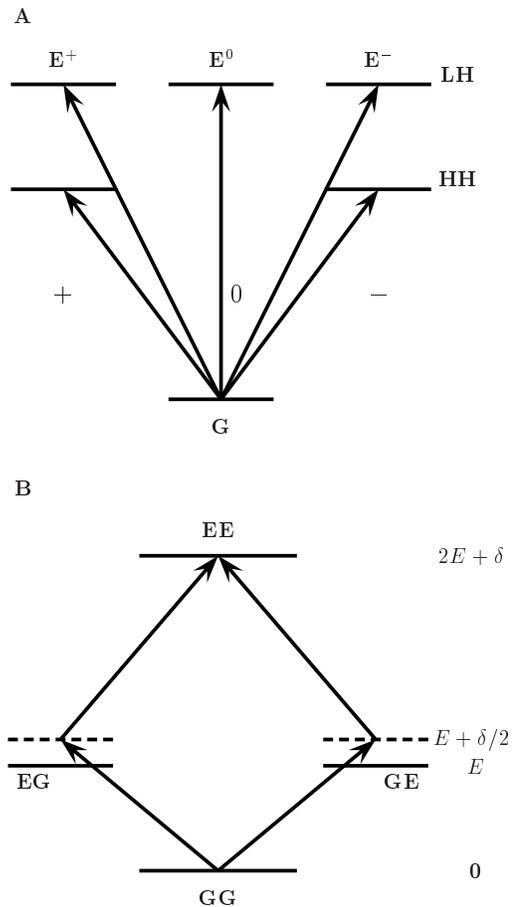}
  \caption{\label{fig1}
    Schematic diagram of the energy-level structure of LH and HH valence-band states (A) and of a typical two-photon process
    (B) in GaAs-based semiconductor macroatoms.}
\end{figure}

A properly tailored laser excitation may promote electrons from
the valence to the conduction band in an energy-selective fashion \cite{26}.
For  the HH the only allowed transitions are
$
|\frac{3}{2}, \frac{3}{2} \rangle \rightarrow |\frac{1}{2}, \frac{1}{2}\rangle,\;
   | \frac{3}{2}, -\frac{3}{2} \rangle \rightarrow |\frac{1}{2}, -\frac{1}{2}\rangle
$.
Here, the first transition is produced by light with  left circular polarization
(usually referred to as $\sigma^{-}$) while the second transition
is produced by light with right circular polarization ($\sigma^{+}$).
In contrast, due to the different structure of their wavefunctions
for the LH we have more allowed transitions \cite{23};
 As for the HH, we have
$
|\frac{3}{2}, \frac{1}{2}\rangle \rightarrow
 |\frac{1}{2}, -\frac{1}{2}\rangle,\;
|\frac{3}{2}, -\frac{1}{2}\rangle \rightarrow |\frac{1}{2}, \frac{1}{2}\rangle
$.
These transitions may be induced by light propagating along the $z$ direction with circular (left or right) polarization.
Moreover, for light propagating along the $x-y$ plane with polarization
along $z$ ($\sigma^0$)
the following transitions are also allowed (and experimentally observed~\cite{28}:
$
|\frac{3}{2}, \frac{1}{2}\rangle \rightarrow |\frac{1}{2}, \frac{1}{2} \rangle, \;
|\frac{3}{2}, -\frac{1}{2}\rangle \rightarrow
|\frac{1}{2}, -\frac{1}{2}\rangle
$.
As a result, we see that by exciting LH electrons with three different kinds of
light ---left- and right-circular polarization as well as linear polarization along $z$---
we can induce three different transitions with the same energy:
$|G\rangle\mapsto|E^\alpha\rangle,\,(\alpha=\pm,0)$ where $|G\rangle$ denotes the ground state of the semiconductor crystal.
The allowed optical transitions as well as the corresponding
energy-level structure for HH and LH are schematically depicted in
Fig.~\ref{fig1}(A).
For the case of a laser excitation resonant with the three degenerate LH transitions,
the corresponding light-matter interaction Hamiltonian is of the form:
\begin{equation}
  H_{int} =
 - \hbar \sum_{\mu=0,\pm}(  \Omega_{\mu,LH}|E^{\mu}\rangle
   \langle G|+\mbox{h.c.} )
\end{equation}

\begin{figure}[t]
  \includegraphics[height=16cm,width=9.5cm]{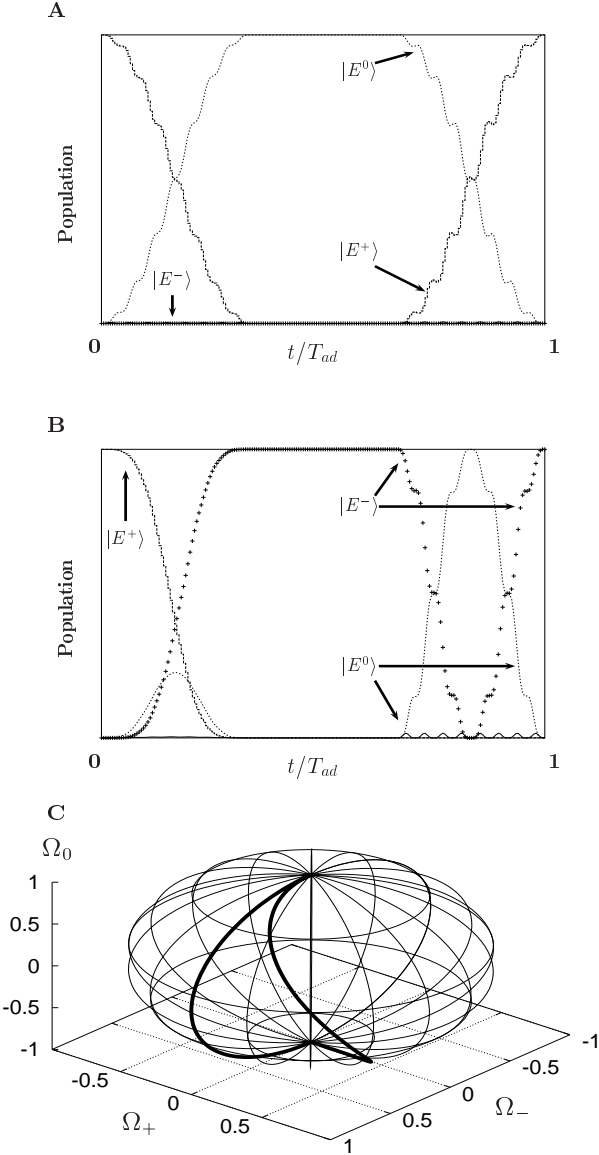}
  \caption{\label{fig2} (A) Simulated time evolution of the HQC gate
    1 with $\phi_1=\pi/4$ and initial state $|E^{+}\rangle$. (B)
    Simulated time evolution of the HQC gate 2 with $\phi_2=\pi/2$ and
    initial state $|E^{+}\rangle$. 
    (C) Simulated quantum evolution of
    gate 2 in the control parameter manifold ($\Omega^{-}$,
    $\Omega^{+}$, $\Omega^{0}$). In these simulated
    experiments we have chosen $\Omega^{-1} = 50$\,fs and $T_{ad} =
    7.5$\,ps (see text).}
\end{figure}

\begin{figure}[t]
  \includegraphics[height=5cm]{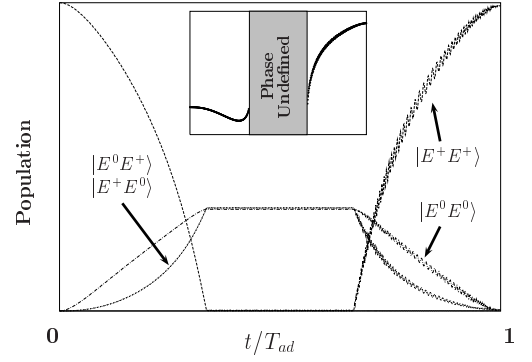}
  \caption{\label{fig3}
    Simulated control-shift over the state $|E^+\rangle^{\otimes\,2}$.
The inset shows (where it is defined) the quantity $ \varphi^+$ where $ \varphi^+ := \mbox{Arg}  \langle\Psi(t)|E^+ E^+\rangle/  |
\langle\Psi(t)|E^+ E^+\rangle|$ 
The values of the parameters are $\delta= 5 \mbox{meV}$, $|\Omega_{0,+}| =\delta/5$, $T_{ad}=.8 \mbox{ns}$;
The gate fidelity $F=|\langle E^+ E^+|\Psi(T_{ad})\rangle|^2=.9899.$}
\end{figure}

This Hamiltonian  has the same structure as the one for trapped-ion internal
levels
analyzed in \cite{18}.
Indeed, for each value of the Rabi couplings $\Omega$'s it admits a couple of {\em dark } states,
i.e., two states $|D_\alpha(\Omega)\rangle \,(\alpha=0,1)$ corresponding to a zero eigenvalue.
These dark states, in a distinguished point in the $\Omega$ space will encode  our qubit.
The quantum manipulations will be realized by the holonomies
$P\exp\oint_\gamma A$
associated to the Wilczeck $u(2)$-valued connection $A$
defined by
$(A_\mu)_{\alpha\beta}=\langle D_\alpha|{\partial}/{\partial\Omega^\mu}| D_\beta\rangle
\quad(\alpha,\beta=0,1;\,\mu=0,\pm)
$.
Our computational basis is given by
$|1\rangle:= |E^{+}\rangle$ and $|0\rangle:= |E^{-}\rangle$.
The state $|E^{0}\rangle$ will play the role of an {\it ancilla},
used, as an auxiliary resource.

To achieve single-qubit universality  is sufficient to enact a couple
of non-commuting single-qubit gates $U_1$ and $U_2$ \cite{22}.
Following Ref.~\cite{18},
for the first gate we choose $\Omega_{-} = 0$,
$\Omega_{+} = -\Omega \sin(\theta /2) ~e^{i \varphi}$ and
$\Omega_{0} = \Omega \cos(\theta /2)$. The dark states
are given by  $|E^{-}\rangle$ and
$|\psi\rangle= \cos(\theta /2) |E^{+}\rangle + \sin(\theta /2)~ e^{i \varphi} |E^{0}\rangle
$.
By evaluating the connection  associated to this
two-dimensional degenerate eigenspace, it
is not difficult to see that    the unitary transformation
$U_1 = e^{i \phi_1 |E^{+}\rangle\langle E^{+}|}$
($\phi_1 =\frac{1}{2} \oint \sin\theta~ d\theta~d\psi$) can be realized as an
holonomy.
For the second gate we choose $\Omega_{-} = \Omega \sin\theta \cos\varphi$,
$\Omega_{+} = \Omega \sin\theta \sin\varphi$ and
$\Omega_{0} = \Omega \cos\theta$. The dark states are now given by
$|\psi_1\rangle = \cos\theta \cos\varphi |E^{-}\rangle + \cos\theta \sin\varphi |E^{+}\rangle - \sin\theta |E^{0}\rangle$ and
$|\psi_2\rangle = \cos\varphi |E^{+}\rangle - \sin\varphi |E^{-}\rangle$.
In this case, the unitary transformation
$U_2 = e^{i \phi_2 \sigma_y}$ where $\phi_2 = \oint \sin\theta d\theta
d\psi$ can be implemented.

For the implementation of the two-qubit gate  we resort to  the
exciton-exciton dipole coupling in semiconductor macromolecules proposed 
in~\cite{24}.
Indeed, if we have two Coulomb coupled quantum dots the presence  of an exciton in
one of them (e.g., in dot b) produces a shift in the energy level of the other one
(e.g., dot a) from $E$ to $E +  \delta$; the total energy in the process is
$2 E+\delta$.
Let us consider the two dots in the ground state $|G G\rangle$;
if we shine them with
light resonant with $E+\delta/2$, we should be able to produce two
excitons $|E E\rangle$.
This is a second-order --- two-photon--- process, i.e., it involves
a virtual transition to
the intermediate states $|EG\rangle$ and $|GE\rangle$ [see Fig.~\ref{fig1}(B)].
Due to energy conservation this is the only possible transition
(the first-order ---or single-photon--- absorption is at energy $E$).
Using different polarizations  $(\sigma_+,\sigma_-,\sigma_0)$ all the
degenerate second-order transitions $|GG  \rangle \rightarrow |E^{\alpha}E^{\beta}\rangle,
\,(\alpha,\beta =0,+,-) $ can be excited. 

This process may be described by the following (effective) two photon Hamiltonian
\begin{eqnarray}
   H_{int} & = & -\frac{2 \hbar^2}{\delta} \sum_{\alpha,\beta =0,+}
   (\Omega_{\alpha} \Omega_{\beta}   |E^{\alpha}, E^{\beta}\rangle \langle G, G|  +
\mbox{h.c.})\ ,
   \label{eq:two_ph}
\end{eqnarray}
where $\Omega_{+,0}$ is the Rabi frequency for the
single-photon process  within second order perturbation
theory. 
Here we have a three-dimensional dark-state manifold; by  evaluating the
the assocaited $u(3)$-valued connection form one can check in a straightforward way that universal control is this dark space 
can be achieved  in a fully holonomic fashion \cite{sol2}. An explicit result will be shown later on.

To test the viability of the proposed HQC implementation scheme in state-of-the-art semiconductor 
nanostructures,
we have performed a direct time-dependent
simulation of gate 1 as well as gate 2. 
To this end, we have chosen $\Omega^{-1} = 50$\,fs 
and as evolution time
$T_{ad} = 7.5$\ ,ps to satisfy adiabaticity.
Moreover, We have choosen as initial state $|\psi(0)\rangle = |E^{+}\rangle$, and the loop such to have $2\phi_1=\phi_2 = \pi /2$ .
The computational states at the end of our adiabatic loop are
$U_1 |E^{+}\rangle= \exp(i \frac{\pi}{4}|E^{+}\rangle\langle E^{+}|)
|E^{+}\rangle = \frac{1+i}{\sqrt{2}}|E^{+}\rangle$ for gate 1
 and  $U_2 |E^{+}\rangle = \exp(i \frac{\pi}{2}\sigma_y) |E^{+}\rangle = |E^{-}\rangle$ for gate 2.
Figure \ref{fig2} shows the state populations
during the quantum-mechanical evolution; as we can see, the state $|G\rangle$
is never populated (as expected in the adiabatic limit).
For the case of gate 1 [see Fig.~\ref{fig2}(a)]
the $|E^{-}\rangle$ state is decoupled in the evolution while
the state $|E^{+}\rangle$ evolves to the {\it ancilla} state ($|E^{0}\rangle$), to eventually end in
$|E^{+}\rangle$ (as we expect for the dark state).
For the case of gate 2 [Fig.~\ref{fig2}(b)] the initial state $|E^+\rangle$
evolves in $|E^-\rangle$ then in $|E^0\rangle$ to end in $|E^-\rangle$; so 
we apply a $Not$ gate.
In Fig.~\ref{fig2}(c) is shown the loop in the control parameters manifold 
($\Omega^{-}$, $\Omega^{+}$, $\Omega^{0}$) for gate 2. 

We also performed a time-dependent simulation of a two-qubit gate, the effective hamiltonian (\ref{eq:two_ph}) has been used.
Figure (\ref{fig3})  shows how a controlled-phase shift over the state $|E^+\rangle^{\otimes\,2}$ can be realized.  
It is important to notice here that the adiabaticity requirement along
with the condition necessary for the validity of a second-order perturbative approximation implies
that $ T_{ad}\gg \delta/|\Omega_{0,+}|^2\gg  1/|\Omega_{0,+}|.$ This means that the operation time
for the two-qubit gates are necessarily longer than the ones for the single-qubit.
In view of the fast dephasing times in  excitonic system this latter fact would result
in a lack of operation fidelity; this drawback has to be mitigated by  a careful parameter optimization.

The simulated experiments in Fig.~\ref{fig2} clearly show that the proposed HQC implementation scheme is fully compatible with realistic parameters of 
state-of-the-art semiconductor nanostructures~\cite{29} as well as with current ultrafast laser technology \cite{26}, prerequisite for its concrete 
realization. Indeed, our simulation shows that (i) one is able to work in the adiabatic limit, and (ii)
our all-optical scheme allows for picosecond gating times; the ``ultralong'' exciton dephasing (on the nanosecond time-scale) recently measured in 
state-of-the-art QD structures~\cite{30} indicate that within the proposed HQC implementation scheme one should be able to perform a few operations
within the dephasing time. In this respect let us stress that our aim here is not to achieve the error rate threshold for massive
fault-tolerant QIP rather to  to demonstrate how highly non-trivial non-abelian quantum phases can be used to realize elementary 
quantum state manipulations  in a semiconductor based-nanostructures    
 
In summary, we have proposed the first implementation scheme for the realization of
non-Abelian  geometrical gates in semiconductor nanostructures.
Our quantum hardware consists of state-of-the-art Coulomb-coupled semiconductor macroatoms; quantum bits are encoded in the dark 
states of polarization-selective excitonic transitions, driven by ultrafast  laser pulses;
the key ingredient for the implementation of the proposed two-qubit gate is
dipole-dipole coupling between excitons in neighboring quantum dots.
The proposed scheme 
 combines the benefits of geometrical QIP 
with the distinguished characteristics of
all-optical implementations in nanostructured semiconductors.


\begin{thebibliography}{99}

\bibitem{1}
For a review see D.P. Di Vincenzo and C.H. Bennett 
{\it Nature} {\bf 404}, 247-255 (2000).

\bibitem{2}
E.Knill and R. Laflamme,  
{\it Phys. Rev.A} {\bf 55}, 900 (1997) and references therein.

\bibitem{3}
P. Zanardi and M. Rasetti,  
{\it Phys. Rev. Lett.} {\bf 79}, 3306 (1997).

\bibitem{4}
L. Viola, E. Knill and S. Lloyd,
{\it Phys. Rev. Lett.} {\bf 82}, 2417 (1999) and references therein.

\bibitem{5} 
 P. Zanardi,  {\it Phys. Lett.A}
{\bf 258}, 77 (1999).

\bibitem{6}
A. Kitaev, 
Preprint (available at http://arXiv.org/abs/quant-ph/9707021).

\bibitem{7} 
M.H. Freedman, M. Larsen and W.  Zhenghan,
Preprint quant-ph/0001108.

\bibitem{8}
P. Zanardi and M. Rasetti,
{\it Phys. Lett.A} {\bf 264}, 94 (1999).

\bibitem{9} 
J. Pachos, P. Zanardi and M. Rasetti, 
{\it Phys. Rev.A} {\bf 61}, 010305(R) (2000).

\bibitem{10}
A. Shapere and F. Wilczek, Eds., {\it Geometric Phases in 
Physics}. (World Scientific, 1989).

\bibitem{11} F. Wilczek and A. Zee,
{\it Phys. Rev. Lett.} {\bf 52}, 2111 (1984).

\bibitem{12} 
J.A. Jones, V. Vedral, A. Ekert, G. Castagnoli, 
 {\it Nature}  {\bf 403}, 869 (2000).

\bibitem{13} G. Falci, R. Fazio, G. M. Palma, J. Siewert, V. Vedral, 
{\it Nature}  {\bf 407}, 355 (2000).

\bibitem{14} 
B.  Paredes, P. Fedichev, J.I. Cirac, P.  Zoller,  
{\it Phys. Rev.Lett.} 87, 010402 (2001).

\bibitem{15} 
 W. Xiang-Bin and  M. Keiji, 
{\it Phys. Rev. Lett.} {\bf 87},  097901 (2001).

\bibitem{16} X.-Q. Li, L.-X. Cen,  G.-X. Huang, L. Ma, Y. Yan, 
quant-ph/0204028. 

\bibitem{17}  
R.G. Unanyan, B.W.  Shore and  K. Bergmann,
{\it Phys. Rev. A} {\bf 59}, 2910 (1999). 

\bibitem{18} 
L.-M. Duan, J. I.  Cirac and P. Zoller,
{\it Science} {\bf 292}, 1695 (2001). 

\bibitem{19} 
L. Faoro, J. Siewert and R. Fazio,
cond-mat/0202217. 

\bibitem{20}
I. Fuentes-Guridi, J. Pachos, S. Bose, V. Vedral, S. Choi,
Phys. Rev. A {\bf 66}, 022102 (2002)

\bibitem{21} 
A. Recati, T. Calarco, P. Zanardi, J. I. Cirac, P. Zoller, 
Phys. Rev. A {\bf 66}, 0302309 (2002)

\bibitem{22} D. Deutsch,A. Barenco and A.  Ekert,
{\it Proc. R. Soc. London A}, {\bf 449}, 669 (1995).

\bibitem{23} 
G. Bastard,  {\it Wave mechanics applied to semiconductor
        heterestructures}. (les Editions De Physique, France,1988). 

\bibitem{24} 
E. Biolatti, R. C. Iotti, P. Zanardi, F. Fausto Rossi,
{ Phys. Rev. Lett} {\bf 85}, 5647 (2000); 
E. Biolatti, I. D'Amico, P. Zanardi, and F. Rossi,
Phys. Rev. B {\bf 65}, 075306 (2002)

\bibitem{25}
L. Jacak,P. Hawrylak and A. Wojs,  {\it Quantum Dots}(Springer, Berlin, 1998).

\bibitem{26}
J. Shah, {\it Ultrafast Spectroscopy of Semiconductors and Semiconductor
Nanostructures} (Springer, Berlin, 1996).

\bibitem{27} J. Preskill, in {\it Introduction to Quantum Computation
        and Information}. H.-K. Lo,S.  Poposcu, T.Spiller, Eds.
        (World Scientific, Singapore, 1999). 

\bibitem{28} 
J.-Y. Marzin, M.N. Charasse and  B. Sermage, 
{\it  Phys. Rev. B} {\bf 31},8298 (1985).

\bibitem{29}
M. Bayer, P. Hawrylak, K. Hinzer, S. Fafard, M. Korkusinski, 
Z. R. Wasilewski, O. Stern, and A. Forchel {\it Science} {\bf 291}, 451 (2001)

\bibitem{30}
P. Borri, W. Langbein, S. Schneider, and U. Woggon
{\it Phys. Rev. Lett.} {\bf 87}, 157401 (2001).

\bibitem{sol2} P. Solinas, P. Zanardi, N. Zangh\`{\i}, F. Rossi
        to be pubblished.

\end{thebibliography}
\end{document}